\documentclass{iau_FM}
\usepackage{graphicx}

\title[Cosmic Radio Dipole] 
{SKA and the Cosmic Radio Dipole} 

\author[Dominik J. Schwarz et al.]   
{Dominik J.~Schwarz$^{1*}$,
Carlos A.~P.~Bengaly$^2$, 
Roy Maartens$^{2,3}$,
\and Thilo M.Siewert$^1$}

\affiliation{$^1$Fakult\"at f\"ur Physik, Universit\"at Bielefeld,
Postfach 100131, 33501 Bielefeld, Germany \\ 
$^*$email: {\tt dschwarz@physik.uni-bielefeld.de} \\[\affilskip]
$^2$Department of Physics \& Astronomy, University of the Western Cape, Cape Town 7535, South Africa\\
$^3$Institute of Cosmology \& Gravitation, University of Portsmouth, Portsmouth PO1 3FX, United Kingdom}

\pubyear{2018}
\jname{Astronomy in Focus, Volume 1} 
\editors{Volker Beckmann \& Claudio Ricci, eds.}
\begin{document}
\maketitle

\begin{abstract}
We study the prospects to measure the cosmic radio dipole by means of continuum surveys with the Square Kilometre Array.  
Such a measurement will allow a critical test of the cosmological principle. It will test whether the 
cosmic rest frame defined by the cosmic microwave background at photon decoupling agrees with the cosmic rest frame of 
matter at late times.  

\end{abstract}


The Square Kilometer Array (SKA) will enable the observation of extragalactic radio sources up to large cosmological distances. 
The SKA's high survey speed offers the unique possibility to survey wide areas 
and to probe the largest observationally available scales in the Universe. The largest feature on the radio sky 
is the cosmic radio dipole. 

The standard model of cosmology predicts that the radio sky must be isotropic. 
Deviations from isotropy are expected to arise from the proper motion of the Solar system w.r.t.\ to the 
isotropic sky. This kinematic signal is expected to be contaminated by a dipole from the matter distribution in the 
local large scale structure and from light propagation effects. We show that the SKA will be able to measure the kinematic dipole. 

It is expected that this kinematic radio dipole agrees with the cosmic microwave background (CMB) dipole, 
which is assumed to be caused by the proper motion of the Sun with respect to the cosmic 
heat bath. The CMB dipole establishes the frame of the comoving observers, 
a central concept in modern cosmology, i.e. the observers that are at rest with respect to a 
spatially flat Friedmann-Lema\^itre space-time. 

The CMB dipole has been measured by Planck with high accuracy and allows us to infer a proper 
motion of the Sun with a speed of $v = (369.82 \pm 0.11)$ km/s towards the Galactic coordinates 
$l = (264.021 \pm 0.011)^\circ$ and $b = (48.253 \pm 0.005)^\circ$ (\cite{PlanckDipole}). However, the CMB dipole 
could also contain other contributions, e.g. a primordial temperature dipole or an integrated 
Sachs-Wolfe effect. Both effects are expected to be sub-dominant, but the cosmic variance 
of the dipole is large.

The extragalactic radio sky offers an excellent opportunity to perform an independent test of the proper 
motion hypothesis. It is expected that the radio dipole is dominated by the kinematic 
dipole, similar to the CMB. This is not the case for galaxy surveys at visible or infrared wavebands, which probe 
much lower redshifts. 

The kinematic dipole in the radio source counts is due to Doppler and aberration 
effects and leads to a dipole amplitude of~\cite{Ellis:1984}
\begin{equation}
\label{eq:A_kin}
A = [2 + x(1 + \alpha)]\beta \;,
\end{equation}
assuming a universal power-spectrum of the flux density $S \propto \nu^{- \alpha}$, 
a scaling of number counts according to the relation $N(>S) \propto S^{-x}$, and $\beta $ the 
dimensionless velocity of our motion relative to the radio sky. We will assume hereafter 
$\alpha = 0.76$, $x=1$ and $\beta \equiv v/c$, hence giving a fiducial
kinematic dipole amplitude of $A = 0.0046$. 

\begin{figure}
\includegraphics[width=0.5\linewidth]{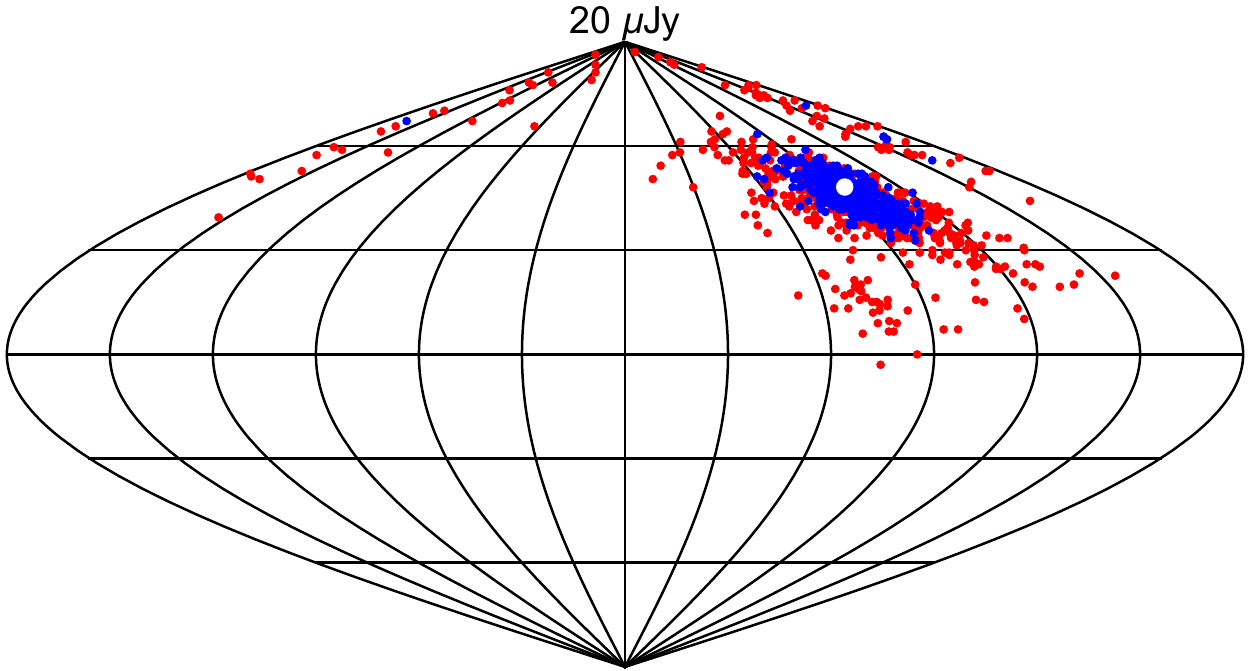}
\includegraphics[width=0.5\linewidth]{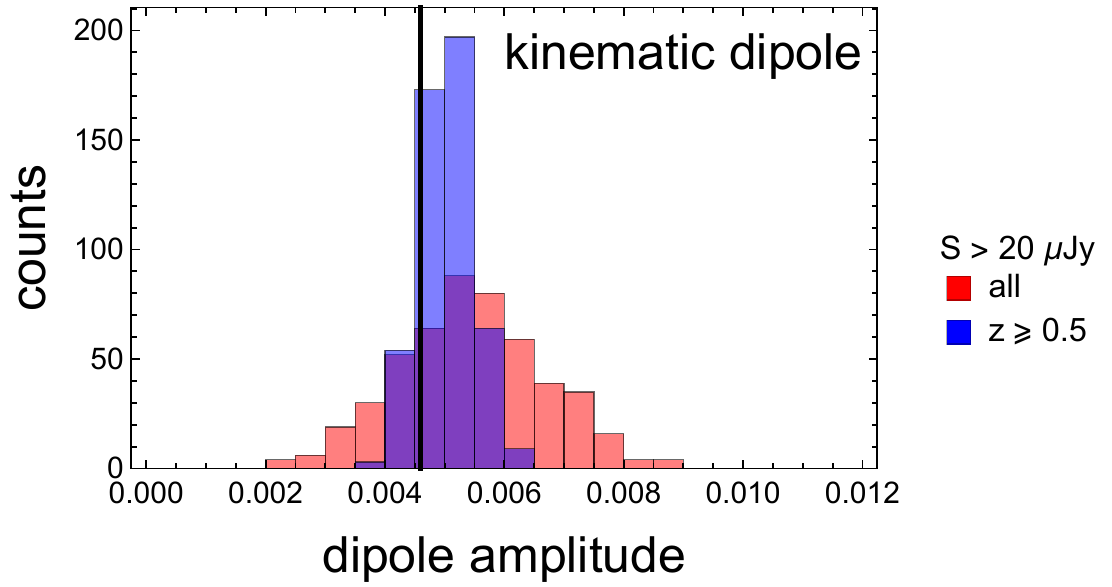}
\caption{Reconstructed radio dipole direction (left) and amplitude (right) for 500 simulations of an SKA1 wide area 
continuum survey in frequency band 1. The red points include all radio sources, for the blue points we assume that local 
sources (at redshift below $0.5$) can be identified and excluded from the analysis. The direction and amplitude of the kinematic 
radio dipole show good agreement with the assumed fiducial values (white point and vertical line). \label{fig}}
\end{figure}

In order to test the ability to measure the cosmic radio dipole, we create mock catalogues on which we 
test our dipole estimators. The mock catalogues and our analysis is described in detail in \cite{Bengaly}. 
Here we present results for a wide area survey with SKA phase 1 by means of the SKA-mid frequency instrument 
in band 1 (350~MHz to 1050MHz) and 
assume a continuum flux density limit of $S > 20 \mu$Jy (see Fig.\ref{fig}).  

The result for the estimated radio dipole from 500 simulations is 
$l = (260 \pm 20)^\circ$ and $b = (45 \pm 16)^\circ$, when all radio sources are taken into account. The significant scatter 
is due to the contribution of the local matter dipole, i.e.\ the inhomogenous distribution of radio objects at small redshifts. This scatter 
can be reduced when local sources are excluded from the analysis. We assume that we will be able to identify the objects below 
redshift of $0.5$ and exclude them from the analysis. That is likely to be possible by means of 
redshift measurements based on an SKA HI survey or based on photometric redshifts from surveys in the infrared and visible.
Excluding local structures leads to a slight increase of Poisson noise, 
but the random admixture of the local structure dipole is largely suppressed. The resulting radio dipole points 
towards  $l = (264.9 \pm 5.6)^\circ$ and $b = (46.7 \pm 5.1)^\circ$, in excellent agreement with the CMB dipole direction. 
Similar improvement is seen in the dipole amplitiude, which reduces from $A = 0.0052 \pm 0.0012$ 
to $A = 0.0048 \pm 0.0005$, when local radio sources are excluded. The estimate is in excellent agreement with the fiducial 
value, with an uncertainty on the kinematic radio dipole of $10$ per cent.

This analysis goes beyond and complements the results presented in \cite{Schwarz:2015pqa}, where the effects of Poisson noise, 
survey area, kinematic dipole and declination dependent systematics have been studied. We now included the effects of 
large scale structure and relativistic light propagation effects.  
\bigskip

\small{CB and RM acknowledge support from the South African SKA Project and the National Research Foundation of South Africa (Grant No. 75415). RM was also supported by the UK Science \& Technology Facilities Council (Grant No. ST/N000668/1). TMS and DJS acknowledge support from DFG within project RTG 1620 ``Models of Gravity''.}

\end{document}